\documentclass[11pt]{article}
\usepackage[totalwidth=13.0cm,totalheight=20.0cm]{geometry}
\usepackage{latexsym,amsthm,amsmath,amssymb,url}

\usepackage{times}
\usepackage{soul}
\usepackage[hidelinks]{hyperref}
\usepackage[utf8]{inputenc}
\usepackage[small]{caption}
\usepackage{graphicx}
\usepackage{sgame}
\usepackage{amsthm}
\usepackage{enumitem}
\usepackage{booktabs}
\usepackage{algorithm}
\usepackage{algorithmic}
\usepackage[switch]{lineno}

\usepackage{amsfonts}
\usepackage{mathrsfs}
\usepackage{amssymb}
\usepackage{tikz}
\usetikzlibrary{arrows}
\usetikzlibrary{arrows.meta}
\usepackage{multirow}
\usepackage{graphics}
\usepackage{xspace}
\usepackage{comment}

\newtheorem{theorem}{Theorem}
\newtheorem{corollary}[theorem]{Corollary}
\newtheorem{lemma}[theorem]{Lemma}
\newtheorem{definition}[theorem]{Definition}
\newtheorem{proposition}[theorem]{Proposition}

\newcommand{\NP}{\textsf{NP}}

\newcommand{\maxweightpartition}{\textsc{Maximum Weighted Digraph Partition}\xspace}
\newcommand{\mwdp}{\textsc{MWDP}\xspace}
\newcommand{\maxweightorpartition}{\textsc{Maximum Weighted Orgraph Partition}\xspace}
\newcommand{\mwop}{\textsc{MWOP}\xspace}
\newcommand{\maxweightsdpartition}{\textsc{Maximum Weighted Symmetric Digraph Partition}\xspace}
\newcommand{\mwsdp}{\textsc{MWSDP}\xspace}

\newcommand{\one}{\textsf{one}\xspace}
\newcommand{\two}{\textsf{two}\xspace}

\newcommand{\sbf}{\ensuremath{\mathbf{s}}\xspace}

\newcommand{\wcal}{\ensuremath{\mathcal{W}}\xspace}
\newcommand{\fcal}{\ensuremath{\mathcal{F}}\xspace}

\newcommand{\PN}[1]{{#1}}

\newcommand{\GG}[1]{{#1}}

\newcommand{\pf}{\textbf{Proof:} }
\newcommand{\2}{\vspace{0.2cm}}

\title{Complexity Dichotomies for the Maximum Weighted Digraph Partition Problem}

\author{
Argyrios Deligkas\thanks{Department of Computer Science, Royal Holloway, University of London, UK, {\tt argyrios.deligkas@rhul.ac.uk}}
\hspace{2mm}
Eduard Eiben\thanks{Department of Computer Science, Royal Holloway, University of London, UK, {\tt eduard.eiben@rhul.ac.uk}}
\hspace{2mm}
Gregory Gutin\thanks{Department of Computer Science, Royal Holloway, University of London, UK, {\tt g.gutin@rhul.ac.uk}}
\hspace{2mm}
Philip R.\ Neary\thanks{Department of Economics, Royal Holloway, University of London, UK, {\tt philip.neary@rhul.ac.uk}}
\hspace{2mm}
Anders Yeo\thanks{Department of Mathematics and Computer Science (IMADA), Southern University of Denmark, Denmark, {\tt   andersyeo@gmail.com   }}
}

\begin{document}

\maketitle

\begin{abstract}
\noindent
\GG{We introduce and study a new optimization problem on digraphs, termed Maximum Weighted Digraph Partition (MWDP) problem. 
We prove three complexity dichotomies for MWDP:\ on arbitrary digraphs, on oriented digraphs, and on symmetric digraphs. 
We demonstrate applications of the dichotomies for binary-action polymatrix games and several graph theory problems. }
\end{abstract}

\section{Introduction}
\label{sec:intro}

\GG{This paper introduces a new optimization problem on digraphs that we term \maxweightpartition problem (\mwdp). We consider its computational complexity on arbitrary digraphs, on {\em oriented digraphs} (or, {\em orgraphs}), in which no arc $xy$ has the {\em opposite arc} $yx$, and on {\em symmetric digraphs}, in which every arc has the opposite arc; terminology and notation not given in this paper can be found in \cite{Bang-JensenG18}.  We prove complexity dichotomies for each of these classes of digraphs.
Somewhat surprisingly, the dichotomies on arbitrary digraphs, on oriented graphs and on symmetric digraphs actually coincide. The common dichotomy is stated in this section, but 
its proof is quite technical and lengthy and is given in Sections \ref{sec:proof1} and \ref{sec:proof2}.
The motivation for this project was binary-action polymatrix games, and an application of our dichotomy on symmetric digraphs to binary-action polymatrix games in Section \ref{sec:games}.\footnote{\PN{Polymatrix games \cite{Janovskaya:1968:} are the most fundamental type of a graphical game \cite{KearnsLittman:2013:C}, that played an important role in resolving the complexity of the Nash equilibrium solution concept \cite{CDT,DGP}.}}
A much more detailed discussion of binary-action polymatrix games and how our complexity dichotomies apply can be found in a companion paper, \cite{Deligkas2nd}. In Section \ref{sec:GTapps}, we discuss applications of our dichotomy for \mwdp on arbitrary digraphs
to well-known and new optimization problems on digraphs. 
A brief discussion completes the paper in Section \ref{sec:disc}, where some open problems are posed. }

\vspace{2mm}
\GG{We will now formulate \mwdp.} 

\paragraph{\bf $\maxweightpartition(\fcal)$.}
Given a family \fcal of \GG{real} $2 \times 2$ matrices, an instance of $\mwdp(\fcal)$ is given by a tuple $(D, c, f)$, where:
\begin{itemize}
    \item $D$ is a \GG{digraph} on $n$ vertices;
    \item $c: A(D) \rightarrow \mathbb{R}^+$ assigns \GG{nonnegative} weights to arcs; 
    \item $f: A(D) \rightarrow \fcal$, is an assignment of a matrix from the family of matrices $\fcal$ to each arc in $D$.
\end{itemize}
Given a partition, $P=(X_1,X_2)$ of $V(D)$, the weight of an arc $uv \in A(D)$ is defined as follows, where 
$M = \left[\begin{array}{cc}
m_{11} & m_{12} \\
m_{21} & m_{22} \\
\end{array} \right]= f(uv)$
is the matrix in $\fcal$
assigned to the arc $uv$.
\begin{equation*}
w^P(uv) = \left\{
\begin{array}{rcl}
c(uv) \cdot m_{11} & \mbox{if} & u,v \in X_1 \\
c(uv) \cdot m_{22} & \mbox{if} & u,v \in X_2 \\
c(uv) \cdot m_{12} & \mbox{if} & u \in X_1 \mbox { and } v \in X_2 \\
c(uv) \cdot m_{21} & \mbox{if} & u \in X_2 \mbox { and } v \in X_1. \\
\end{array}
\right.
\end{equation*}
Given a partition $P$, the weight of $D$, denoted by $w^P(D)$, is defined as the sum of the weights on every arc $uv$.
That is, $w^P(D) = \sum_{a \in A(D)} w^P(a)$.  The goal is to find a partition $P$ that maximizes $w^P(D)$.

We would like to highlight that the direction of an arc in an instance of $\mwdp(\fcal)$ is required to determine which vertex defines the row and which vertex defines the column of the assigned matrix. 

\vspace{2mm}

\GG{
The \maxweightorpartition problem (\mwop) is \mwdp restricted to oriented graphs. 
The \maxweightsdpartition problem (\mwsdp) is \mwdp restricted to symmetriic graphs.
}



\vspace{2mm}

Our main result shows that the tractability of solving $\mwdp(\fcal)$, \GG{$\mwop(\fcal)$ and $\mwsdp(\fcal)$} depends on properties of the matrices in the family of matrices $\fcal$.
We now introduce three properties that a matrix $M \in \fcal$ may satisfy. 
\begin{itemize}
    \item {\bf Property (a):} $m_{11} + m_{22} \geq m_{12} + m_{21}$.
    \item {\bf Property (b):} $m_{11} = \max\{ m_{11}, m_{22}, m_{12}, m_{21} \}$.
    \item {\bf Property (c):} $m_{22} = \max\{ m_{11}, m_{22}, m_{12}, m_{21} \}$.
\end{itemize}

Using the three properties above, we present a common dichotomy for the complexity of $\mwdp(\fcal)$, \GG{$\mwop(\fcal)$ and $\mwsdp(\fcal)$} with respect to \fcal.

\begin{theorem}
\label{thm:dichotomy}
An instance $(D,c, f)$ of $\mwdp(\fcal)$ \GG{($\mwop(\fcal)$, $\mwsdp(\fcal)$, respectively)} can be solved in polynomial time if one of the following holds.
\begin{enumerate}
    \item \label{main-case1} All matrices in \fcal satisfy Property (a).
    \item \label{main-case2} All matrices in \fcal satisfy Property (b).
    \item \label{main-case3} All matrices in \fcal satisfy Property (c).
\end{enumerate}
In every other case, $\mwdp(\fcal)$ \GG{($\mwop(\fcal)$, $\mwsdp(\fcal)$, respectively)} is \NP-hard.
\end{theorem}

Fig. \ref{fig:Ex} illustrates Theorem \ref{thm:dichotomy} and some notions introduced before the theorem.

\newcommand{\Matrix}[4]{\left[ \begin{array}{cc}
#1 & #2 \\
#3 & #4 \\ \end{array} \right]}

\begin{figure}[t]
\begin{center}
\tikzstyle{vertexY}=[circle,draw, top color=gray!5, bottom color=gray!30, minimum size=12pt, scale=0.69, inner sep=0.25pt]
\begin{tikzpicture}[scale=0.20]
\node (x) at (1.0,4.0) [vertexY] {x};
\node (y) at (9.0,4.0) [vertexY] {y};
\node (z) at (17.0,4.0) [vertexY] {z};

\draw [->,line width=0.02cm] (x) -- (y);
\draw [->,line width=0.02cm] (y) -- (z);

\draw (5,7) node {{\tiny $\Matrix{4}{2}{2}{6}$}}; 
\draw (13,7) node {{\tiny $\Matrix{7}{5}{6}{2}$}};

\node at (5,2.8) {{\tiny $c(xy)=1$}};
\node at (13,2.8) {{\tiny $c(yz)=2$}};

\draw [->,dotted, line width=0.02cm] (5,6) -- (5,4.3);
\draw [->,dotted, line width=0.02cm] (13,6) -- (13,4.3);

\node at (9,-3) {(i)};
\node at (39,-3) {(ii)};

\node (xx) at (31.0,4.0) [vertexY] {x};
\node (yy) at (39.0,4.0) [vertexY] {y};
\node (zz) at (47.0,4.0) [vertexY] {z};

\draw[->, line width=0.02cm] (xx) to [out=20,in=160] (yy);
\draw[->, line width=0.02cm] (yy) to [out=20,in=160] (zz);
\draw[->, line width=0.02cm] (zz) to [out=200,in=340] (yy);
\draw[->, line width=0.02cm] (yy) to [out=200,in=340] (xx);

\draw (28.5,7.7) node {{\tiny $c(xy)=2$ \&}};
\draw (26.5,0.3) node {{\tiny $c(yx)=0$ \&}};

\draw (49.5,7.7) node {{\tiny \& $c(yz)=1$ }};
\draw (51.5,0.3) node {{\tiny \& $c(zy)=3$ }};

\draw (35,7.7) node {{\tiny $\Matrix{2}{0}{2}{3}$}};
\draw (43,7.7) node {{\tiny $\Matrix{3}{4}{1}{3}$}};

\draw (33,0.3) node {{\tiny $\Matrix{2}{0}{2}{3}$}};
\draw (45,0.3) node {{\tiny $\Matrix{4}{2}{5}{5}$}};

\draw [->,dotted, line width=0.02cm] (35,6.7) -- (35,5.3);
\draw [->,dotted, line width=0.02cm] (43,6.7) -- (43,5.3);

\draw [->,dotted, line width=0.02cm] (33,1.3) -- (35,2.7);
\draw [->,dotted, line width=0.02cm] (45,1.3) -- (43,2.7);

\end{tikzpicture}
\caption{
 In (i) above is an instance of MWOP($\{M_1,M_2\}$), where $M_1=\tiny{\Matrix{4}{2}{2}{6}}$ and $M_2=\tiny{\Matrix{7}{5}{6}{2}}$.
The instance consists of the shown oriented graph and the values $c(xy)=1$, $f(xy)=M_1$,  $c(yz)=2$  and $f(yz)=M_2$.
As $M_1$ satisfies Properties (a) and (c), but not Property (b) and $M_2$ satisfies Property (b), but not Property (a) or (c) we
note that MWOP($\{M_1,M_2\}$) is an \NP-hard problem. \\
In (ii) is an instance of MWDSP($\{R_1,R_2,R_3\}$), where 
$R_1=\tiny{\Matrix{2}{0}{2}{3}}$,
$R_2=\tiny{\Matrix{3}{4}{1}{3}}$ and
$R_3=\tiny{\Matrix{4}{2}{5}{5}}$.
Note that MWDSP($\{R_1,R_2,R_3\}$) is polynomial time solvable as $R_1$, $R_2$ and $R_3$ all satisfy Property~(a).} \label{fig:Ex}
\end{center} \end{figure}

\GG{Since $\mwop(\fcal)$ and  $\mwsdp(\fcal)$ are  subproblems of $\mwdp(\fcal)$, to prove Theorem~\ref{thm:dichotomy} it suffices to show that $\mwdp(\fcal)$ can be solved in polynomial time if either (a) or (b) or (c) holds, and $\mwop(\fcal)$ and  $\mwsdp(\fcal)$ are \NP-hard if none of (a), (b) and (c) hold. 
Theorem~\ref{thm:dichotomy} decomposes the space into four cases, each of varying difficulty.
Cases~\ref{main-case2} and~\ref{main-case3} are immediate, since they admit trivially-optimal solutions: in Case~\ref{main-case2} all vertices belong to $X_1$ and in Case~\ref{main-case3} all vertices belong to $X_2$. 
On the other hand, Case~\ref{main-case1} is far from trivial and it requires a more sophisticated argument that creates an equivalent min-cut instance on an undirected graph (see Theorem~\ref{lem:case-1}), which is known to be solvable polynomial-time~\cite{korte2011combinatorial}. These three cases for $\mwdp(\fcal)$ are considered in detail in Section \ref{sec:proof1}.
Finally, the last case deals with ``every other case'' and shows that the problem becomes intractable.
The proof for $\mwop(\fcal)$ involves a series of intricate subcases and constructions, see the main part Section \ref{sec:proof2} ending with Theorem \ref{NPhMWOP}. We first prove Theorem \ref{NPhMWOP} for the case when $\cal F$ is a singleton
and then for $\cal F$ of arbitrary size. 
The proof for $\mwsdp(\fcal)$ follows from a simple reduction from $\mwop$ to $\mwsdp$, see Theorem \ref{NPhMWSDP} in Section \ref{sec:proof2}.
}

\2

\GG{An extended abstract of both this paper and the companion paper \cite{Deligkas2nd} were published in the proceedings of IJCAI 2023 \cite{DeligkasIJCAI23}. There are a number of differences between \cite{DeligkasIJCAI23} and this paper. In particular, while \cite{DeligkasIJCAI23} has only one complexity dichotomy for MWDP, this paper proves three such dichotomies.
}

\section{Applications of \GG{MWSDP} in Binary-Action Polymatrix Games}\label{sec:games}


\PN{
Polymatrix games, \cite{Janovskaya:1968:}, are easily described.
There is an undirected graph where vertices represent players and edges represent a two-player game played between the adjacent players (vertices).
A player's payoff is the sum of payoffs earned from interacting with every neighbouring player, where the same action must be used with each. 
While simple, polymatrix games have proved to be of fundamental importance.
They are the ``go-to'' way of modeling coordination games on networks~\cite{apt2017coordination,apt2022coordination,Blume:1993:GEB,Ellison:1993:E,Morris:2000:RES,Neary:2012:GEB,NearyNewton:2017:JMID,Peski:2010:JET,rahn2015efficient} and additively separable hedonic games~\cite{bogomolnaia2002stability}, they provide the building-blocks for hardness reductions~\cite{CDT,DGP,DFHM22-focs,Rubinstein18-polymatrix}, and have even been applied to protein-function prediction and semi-supervised learning~\cite{elezi2018transductive-gtg,vascon2020protein-gtg}.

One natural measure of interest in polymatrix games, as indeed it is in all games, is determining the pattern of behaviour that maximises social welfare, i.e., what strategies should each player select so as to maximise the sum of all payoffs.
In a polymatrix game this is easily described.
The welfare of each edge is simply the sum of the payoffs of both players in the outcome of the game on each edge.
Thus, each 2-player ``game matrix'', with two entries in each cell (one payoff for each player) is replaced by a standard matrix with one entry in each cell.
For polymatrix games where each player has only two strategies, it is straightforward to see that the issue of optimal welfare reduces to a particular case of \mwdp.

Let us now be more formal.

An $n$-player binary-action \emph{polymatrix game} is defined by a graph $G$, where each vertex represents a player.}
Each player $i \in V(G)$ has two actions called \one and \two. For each edge $ij \in E(G)$, there is a $2 \times 2$ two-player game $(\Pi^{ij}, \Pi^{ji})$, where matrix $\Pi^{ij}$ gives the {\em payoffs} that player $i$ obtains from their interaction with player $j$, and likewise matrix $\Pi^{ji}$ gives the payoffs  player $j$ gets with the interaction with player $i$. 

A pure \emph{strategy profile} $\sbf = (s_1, s_2, \dots, s_n)$ specifies an action for each of the players; we will use $S$ to denote the set of all strategy profiles.
It is convenient to think of $s_i=(1,0)^{T}$ when player $i$ chooses action \one and $s_i=(0,1)^{T}$ when they choose action \two.
For each strategy profile $\sbf \in S$, the {\em payoff} of player $i$ is 
$p_i(\sbf) := s_i^{T} \cdot \sum_{j \; :\; ij \in E(G)} \Pi^{ij} \cdot s_j$.
In other words, the payoff obtained by a player is the sum of the payoffs obtained from the interaction with every neighboring player, where the same action must be used with each.

We are interested in computing welfare-optimal strategy profiles.

\begin{definition}\label{def:sw}
The {\em social welfare} of strategy profile $\sbf$ is $\wcal{(\sbf{})} := \sum_{i \in V} p_i(\sbf{})$.
\end{definition}

\GG{Consider an instance of the binary-action polymatrix game with graph $G$ and two payoff matrices $(\Pi^{uv}, \Pi^{vu})$ for each $uv \in E(G)$.
It is natural to formulate such an instance by an  instance of {\mwsdp} as follows. Let $D$ be a symmetric digraph obtained from $G$ by replacing every edge $uv \in E(G)$ by the opposite arcs $uv$ and $vu$ such that the payoff matrix $\Pi^{uv}$ is associated with arc $uv$ and  the payoff matrix $\Pi^{vu}$ is associated with arc $vu$.  
Also, set $c(uv)=c(vu)=1$.

In fact, using symmetric digraphs allows us to generalize the game by allowing $c(uv)$ to be an arbitrary non-negative real. (Here $c(uv)$ indicates the relative importance of payoff associated with arc $uv$.)
The formula in Definition \ref{def:sw} remains the same, but that for $p_i(\sbf)$ is replaced with
$p_i(\sbf) := s_i^{T} \cdot \sum_{j \; :\; ij \in E(G)} c(ij)\Pi^{ij} \cdot s_j$. In what follows, we consider the generalized version of the binary-action polymatrix game.}

Introduction of $D$ induces an immediate translation between strategy profiles and partitions. Player $v$ chooses action \one if and only if the corresponding vertex $v$ in $D$ belongs to $X_1$. Conversely, player $v$ chooses action \two if and only if the corresponding vertex $v$ in $D$ belongs to $X_2$. 

For any strategy profile \sbf, we use $P(\sbf)$ to denote the corresponding partition. The following lemma trivially follows from the reduction above. 

\begin{lemma}
\label{lem:sw-to-mwdp}
For every possible strategy profile $\sbf$, it holds that $\wcal(\sbf) = w^{P(\sbf)}(D)$.
\end{lemma}

So, the combination of Lemma~\ref{lem:sw-to-mwdp} and Theorem~\ref{thm:dichotomy}, yields the following result, which is a clean complexity dichotomy for maximizing social welfare in general binary-action polymatrix games. To the best of our knowledge, this is the first dichotomy of this kind.

\begin{theorem}
\label{thm:sw-general}
Consider a binary-action polymatrix game on input \GG{symmetric digraph $D$. Let ${\cal F} = \{\Pi^{uv}:\ uv \in A(D)\}.$}
Finding a strategy profile that maximizes the social welfare can be solved in polynomial time if one of the following holds (where $\Pi^{uv}=\left[\begin{array}{cc}
\pi^{uv}_{11} & \pi^{uv}_{12} \\
\pi^{uv}_{21} & \pi^{uv}_{21} \\
\end{array} \right]$).
\begin{itemize}
   \item $\pi^{uv}_{11} + \pi^{uv}_{22} \geq \pi^{uv}_{12} + \pi^{uv}_{21}$ for every $uv \in A(D)$.
   \item $\pi^{uv}_{11} \geq \max \{\pi^{uv}_{12}, \pi^{uv}_{21}, \pi^{uv}_{22} \}$ for every $uv \in A(D)$.
   \item $\pi^{uv}_{22} \geq \max \{\pi^{uv}_{11}, \pi^{uv}_{12}, \pi^{uv}_{21} \}$ for every $uv \in A(D)$.
\end{itemize}
In every other case, the problem is \NP-hard.
\end{theorem}


\section{Applications of MWDP and MWOP in Graph Theory}\label{sec:GTapps}

\GG{In this section, we consider a selection of graph theory problems and show reductions between these problems and \mwdp. Some of these problems are well-known and some are new. }

\newcommand{\EquivPolyMM}[1]{{{\bf Reduction to MWDP:} #1}}
\newcommand{\EquivPolyMMf}[1]{{{\bf Reduction from MWDP:} #1}}
\newcommand{\EquivPolyMMO}[1]{{{\bf Reduction to MWOP:} #1}}




\subsection{Maximum Average Degree}
\GG{The {\em maximum average degree} ${\rm mad}(G)$ of a graph $G$ is the maximum over all subgraphs of $G,$ of the average degree of the subgraph.
The maximum average degree is an important graph parameter used in proving various graph theoretic results, see e.g. \cite{NadaraS22} and several papers cited in \cite{NadaraS22} on vertex and edge partitions of graphs.
See also \cite{ChenMW23}, where the authors prove two major conjectures on proper orientation number of planar graphs using properties of maximum average degree. 
The maximum average degree of a graph can be computed in polynomial time \cite{PicardQ82,Gold}. Note that the parameter ${\rm mad}(G)/2$ is called the {\em maximum density} of $G$ and it was studied in several papers (see, e.g.,\cite{PicardQ82,Gold}); in fact, 
\cite{PicardQ82,Gold} prove that it is the maximum density that can be computed in polynomial time. 
}

The {\sc Max Average Degree} problem is as follows. Given a graph $G$ and a real number $k$,  decide whether  ${\rm mad}(G)> k$ and if  ${\rm mad}(G)> k$, then
find a vertex set $X \subseteq V(G)$ such that 
$G[X]$ has average degree greater than $k$. 

\EquivPolyMMO{ Let $M_1 = {\small \left[\begin{array}{cc}
k & 0 \\
0 & 0 \\
\end{array}\right]}$ and let  
$M_2 = {\small \left[\begin{array}{cc}
0 & 0 \\
0 & 2 \\
\end{array}\right]}$. 
}

We will now create a digraph $D$ as follows.
Initially let $D$ be any orientation of $G$. Then,
for each vertex $u \in V(G)$ add a new vertex, $v_u$, and the arc $u v_u$ to $D$. 
This defines $D$. Note that $|V(D)|=2|V(G)|$.
Associate $M_1$ with the arc $u v_u$ for all $u \in V(G)$.
Associate $M_2$ with all other arcs of $D$. We set $c(uv)=1$ for all arcs $uv$ in $D.$

\GG{Assume that $P=(X_1,X_2)$ is an arbitrary partition of $V(D)$ and let $w^P(D)$ denote the weight of this partition.
We may assume that $u$ and $v_u$ belong to the same set in $X_1,X_2$ as moving $v_u$ to the same set as $u$ does not decrease $w^P(D)$.
Let $W=V(G) \cap X_2$ and let $e(W,W)$ denote the number of edges in $G[W]$.
Then 
\[
w^P(D) = k |V(G)| - k |W| + 2e(W,W) 
\]
So $w^P(D)>k |V(G)|$ if and only if $2e(W,W)> k |W|$, which is equivalent to $k < \frac{2e(W,W)}{|W|}$. 
As $2e(W,W)/|W| = ( \sum_{u \in W} d_{G[W]}(u) )/|W|$ we note that the average degree in $G[W]$ is greater than 
$k$ if and only if $w^P(D)>k |V(G)|$.
}

So, by the dichotomy for MWOP, the {\sc Max Average Degree} problem is polynomial time solvable.







\subsection{$2$-color-difference} 

\GG{The {\sc $2$-color-difference} problem is as follows. We are given an edge-weighted $2$-edge-colored graph $G$, where $w_i(X)$ denotes the weight of all edges of color $i$
in  a vertex set $G[X]$.
Find a set $X \subseteq V(G)$, which maximizes $w_2(X)-w_1(X)$. To the best of our knowledge, this is a new natural graph theoretical problem.
}

\EquivPolyMMf{ Let $M_1 = {\small \left[\begin{array}{cc}
1 & 0 \\
0 & 0 \\
\end{array}\right]}$ and let
 $M_2 = {\small \left[\begin{array}{cc}
-1 & 0 \\
0 & 0 \\
\end{array}\right]}$. 
 Note that if we do not want negative weights in $M_2$ we can add 1 to all entries without changing the problem (the value of the result just increases by 1 for every 
arc associated with $M_2$).

Consider the following instance of MWDP(\{$M_1$,$M_2$\}): a digraph $D$ in which every arc $uv$ is associated with a cost $c_{uv}$ and a matrix $M^{uv} \in \{M_1,M_2\}$.
We can reduce this instance to an instance $G$
of {\sc $2$-color-difference} as follows. Let $G$ be the underlying graph of $D$ and for every arc $uv \in A(D)$ we color $uv$ with color 2 in $G$ if $M^{uv}=M_1$ and we color
$uv$ with color 1, otherwise. Costs in $D$ become weights in $G$.
Let $P=(X_1,X_2)$ be a partition of $V(D)$.
Then $w^P(D)$ is the sum of all the costs  of arcs in $X_1$ associated with $M_1$ minus the sum of the costs of arcs in $X_1$ associated with $M_2$.
However, this is exactly $w_2(X_1)-w_1(X_1)$ in $G$. Thus, an optimal solution for the {\sc 2-color-difference} problem corresponds to an optimal solution for MWDP(\{$M_1$,$M_2$\}).  Therefore, as our reduction is   polynomial, our dichotomy result shows that the {\sc 2-color-difference} problem is \NP-hard.
}


\subsection{$2$-color-partition} 

The problem is as follows. Given a $2$-edge-colored graph $G$ find a partition $(X_1,X_2)$ of $V(G)$ which maximizes the 
sum of the number of edges in $G[X_1]$ of color one and the number of edges in $G[X_2]$ of color two. \GG{To the best of our knowledge, this is a new natural graph theoretical problem.}

\EquivPolyMMO{ Let $M_1 = {\small \left[\begin{array}{cc}
1 & 0 \\
0 & 0 \\
\end{array}\right]}$ and let
 $M_2 = {\small \left[\begin{array}{cc}
0 & 0 \\
0 & 1 \\
\end{array}\right]}$.  By associating $M_1$ to any orientation of each edge of color one and 
associating $M_2$ to any orientation of each edge of color two we note that 
our dichotomy implies that this problem is polynomial time solvable.

\subsection{\GG{Closeness to Being Balanced Weighted Digraph}}
\GG{For an arc-weighted digraph $D=(V,A,w),$ let $w^+(x)$ be the sum of weights $w(xy)$ of arcs of the form $xy$, $y\in V$, and $w^-(x)$ the sum of weights $w(yx)$ of arcs of the form $yx$, $y\in V.$
We say that $D$ is {\em balanced} if $w^+(x)=w^-(x)$ for all $x\in V.$

The problem is as follows. Given an arc-weighted  digraph $D$, find a partition $(X_1,X_2)$ of $V$ where the difference between the total weight of arcs from $X_1$ to $X_2$ and  the total weight of arcs from $X_2$ to $X_1$ is maximized. Let us denote this maximum by $r^+(D).$ Note that $r^+(D)$ is equal to the sum of $w^+(x)-w^-(x)$ over all $x\in V$ for which $w^+(x) > w^-(x)$, see, e.g., \cite{AiMDC}. Thus, $r^+(D)$ can be computed in polynomial time. 
Note that $r^+(D)=0$ for balanced arc-weighted digraphs.
} 


\EquivPolyMM{ Let $M = {\small \left[\begin{array}{cc}
0 & 1 \\
-1 & 0 \\
\end{array}\right]}$ for every arc or if we do not allow negative values we can add 1 to all entries of $M$ and get 
 $M' = {\small \left[\begin{array}{cc}
1 & 2 \\
0 & 1 \\
\end{array}\right]}$.  \GG{Let $c(uv)=w(uv)$ for every $uv\in A.$}
\GG{This reduction and the dichotomy for MWDP imply that this problem is polynomial time solvable.}
}

\subsection{Maximum Weighted  Directed  Cut}
This well-known problem is as follows. Given an arc-weighted digraph $D=(V,A,w)$, find a partition $(X_1,X_2)$ of $V$ with the maximum total weight of 
arcs from $X_1$ to $X_2$, see, e.g., \cite{AiMDC}.

\EquivPolyMMf{ Let $M = {\small \left[\begin{array}{cc}
0 & 1 \\
0 & 0 \\
\end{array}\right]}$. Given an instance of MWDP($M$) we let
 $w(uv)=c(uv)$ for all $xy\in A$, thereby obtaining an instance of 
 Maximum Weighted Directed Cut.
This reduction and the dichotomy for MWDP imply that Maximum Weighted Directed Cut is \NP-hard.}

\subsection{Directed Minimum $(s,t)$-cut}
This well-known problem is as follows. Given a digraph $D$ with $s,t \in V(D)$, find an $(s,t)$-partition $(X_1,X_2)$ (i.e., $s \in X_1$ and $t \in X_2$) with the fewest number of arcs from $X_1$ to $X_2$.

This is equivalent to finding the largest number of arc-disjoint paths from $s$ to $t$ (by Menger's Theorem).

\EquivPolyMM{ Let $M = {\small \left[\begin{array}{cc}
1 & 0 \\
1 & 1 \\
\end{array}\right]}$ and 
 $S = {\small \left[\begin{array}{cc}
|A(D)| & 0 \\
0 & 0 \\
\end{array}\right]}$ and 
 $T = {\small \left[\begin{array}{cc}
0 & 0 \\
0 & |A(D)| \\
\end{array}\right]}$. All arcs of $D$ get associated with matrix $M$. We then add a new vertex $s'$ and the arc $s's$ which we associate with matrix $S$. We also add a new vertex $t'$ and the arc $tt'$ which we associate with matrix $T$. Let $c(uv)=1$ for all $uv\in A(D)$.
Now the maximum value we can obtain is $3|A(D)|$ minus the size of a minimum $(s,t)$-cut.  
So by the dichotomy for MWDP, this problem is polynomial time solvable.
}

}

\section{Proving the Polynomial Part of the Dichotomy}\label{sec:proof1}

Let us start by recalling \mwdp(${\cal F}$).%


$\mwdp({\cal F})$: We are given a family of $2 \times 2$ matrices ${\cal F}$. 
An instance of {\it \mwdp(${\cal F}$)} consists of a digraph $D$,  
arc-weights $c: A(D) \rightarrow \mathbb{R}^+$ and 
an assignment, $f: A(D) \rightarrow {\cal F}$, of a matrix from ${\cal F}$ to each arc in $D$.

Given a partition, $P=(X_1,X_2)$, of $V(D)$ the weight of an arc $uv \in E(D)$ is defined as follows, where $M = \left[\begin{array}{cc}
m_{11} & m_{12} \\
m_{21} & m_{22} \\
\end{array}\right]=f(uv)$ is the matrix
assigned to the arc $uv$.

 \[
w^P(uv) = \left\{
\begin{array}{rcl}
c(uv) \cdot m_{11} & \mbox{if} & u,v \in X_1 \\
c(uv) \cdot m_{22} & \mbox{if} & u,v \in X_2 \\
c(uv) \cdot m_{12} & \mbox{if} & u \in X_1 \mbox { and } v \in X_2 \\
c(uv) \cdot m_{21} & \mbox{if} & u \in X_2 \mbox { and } v \in X_1 \\
\end{array}
\right.
\]

  The weight of $D$, $w^P(D)$, is defined as the sum of all the weights of the arcs. That is, $w^P(D) = \sum_{a \in A(D)} w^P(a)$. 
  The aim is to find the partition, $P$, that maximizes $w^P(D)$.

\vspace{2mm}

\begin{theorem}\label{lem:case-1}
If $m_{11} + m_{22} \geq m_{12} + m_{21}$ for all matrices $M \in {\cal F}$ then {\it \mwdp(${\cal F}$)} can be solved in polynomial time.
\end{theorem}
\pf
Let $(D,c,f)$ be an instance of {\it \mwdp(${\cal F}$)}.
We will construct a new (undirected) multigraph $H$ with vertex set $V(D) \cup \{s,t\}$ as follows. Let $G(D)$ denote the undirected multigraph obtained from $D$ by removing
orientations of all arcs. 
Initially let $E(H) = E(G(D)) \cup \{su,tu \; | \; u \in V(D) \}$ 
and let all edges in $H$ have weight zero. Now for each arc $uv \in A(D)$ we modify the weight function $w$ of $H$ as follows, where $M=f(uv)$ is the matrix associated
with the arc $uv$.

\begin{itemize}
\item Let $w(uv)=c(uv) \times \frac{m_{11} + m_{22} - m_{12} - m_{21}}{2}$. Note $w(uv)$ is the weight of the undirected edge $uv$ in $H$ associated with the arc $uv$ in $D$ and that this weight is non-negative.
\item Add $c(uv) \times \left( \frac{-m_{22}}{2}\right)$ to $w(su)$.
\item Add $c(uv) \times \left( \frac{-m_{22}}{2}\right)$ to $w(sv)$.
\item Add $c(uv) \times \frac{m_{21}-m_{11}-m_{12}}{2}$ to $w(tu)$.
\item Add $c(uv) \times \frac{m_{12}-m_{11}-m_{21}}{2}$ to $w(tv)$.
\end{itemize}

Let $\theta$ be the smallest possible weight of all edges in $H$ after we completed the above process ($\theta$ may be negative).
Now consider the weight function $w^*$ obtained from $w$ by 
subtracting $\theta$ from all edges incident with $s$ or $t$. That is, $w^*(uv)=w(uv)$ if $\{u,v\} \cap \{s,t\}=\emptyset$ and 
$w^*(uv)=w(uv) - \theta$ otherwise.  
Now we note that all $w^*$-weights in $H$ are non-negative.  We will show that for any $(s,t)$-cut, $(X_1,X_2)$ in $H$ (i.e., $(X_1,X_2)$ partitions 
$V(H)$ and $s \in X_1$ and $t \in X_2$) the $w^*$-weight of the cut is equal to $-w^P(D) - |V(D)|\theta$, where $P$ is the partition 
$(X_1 \setminus \{s\},X_2 \setminus \{t\})$ in $D$.  Therefore a minimum weight cut in $H$ maximizes $w^P(D)$ in $D$.

Let $(X_1,X_2)$ be any $(s,t)$-cut in $H$.  For every $u \in V(D)$ we note that exactly one of the edges $su$ and $ut$ will belong to the cut.
Therefore, we note that the $w^*$-weight of the cut is $|V(D)|\theta$ less than the $w$-weight of the cut.
 It therefore suffices to show that the $w$-weight of the cut is $-w^P(D)$ (where  $P$ is the partition
$(X_1 \setminus \{s\},X_2 \setminus \{t\})$ of $V(D)$).

Let $uv \in A(D)$ be arbitrary and consider the following four possibilities.

\begin{itemize}
\item $u,v \in X_1$. In this case when we considered the arc $uv$ above (when defining the $w$-weights) we added  
$c(uv) \times \frac{m_{21}-m_{11}-m_{12}}{2}$ to $w(tu)$ and $c(uv) \times \frac{m_{12}-m_{11}-m_{21}}{2}$ to $w(tv)$.
So, we added the following amount to the $w$-weight of the  $(s,t)$-cut, $(X_1,X_2)$.
\[
c(uv) \times \left( \frac{m_{21}-m_{11}-m_{12}}{2} + \frac{m_{12}-m_{11}-m_{21}}{2} \right) = -c(uv) \cdot m_{11}
\]
\item $u,v \in X_2$. In this case when we considered the arc $uv$ above (when defining the $w$-weights) we added
$c(uv) \times \left(\frac{-m_{22}}{2}\right)$ to $w(su)$ and $c(uv) \times \left(\frac{-m_{22}}{2}\right)$ to $w(sv)$.
So, we added the following amount to the $w$-weight of the  $(s,t)$-cut, $(X_1,X_2)$.
\[
c(uv) \times \left( \frac{-m_{22}}{2} + \frac{-m_{22}}{2} \right) = -c(uv) \cdot m_{22}
\]
\item $u \in X_1$ and $v \in X_2$. In this case when we considered the arc $uv$ above (when defining the $w$-weights) we 
let $w(uv)=c(uv) \times \frac{m_{11} + m_{22} - m_{12} - m_{21}}{2}$ and added
$c(uv) \times \left( \frac{-m_{22}}{2}\right)$ to $w(sv)$ and
$c(uv) \times \frac{m_{21}-m_{11}-m_{12}}{2}$ to $w(tu)$.
So, we added the following amount to the $w$-weight of the  $(s,t)$-cut, $(X_1,X_2)$.
\begin{eqnarray*}
c(uv) \times \left( \frac{m_{11} + m_{22} - m_{12} - m_{21}}{2} + \frac{-m_{22}}{2} + \frac{m_{21}-m_{11}-m_{12}}{2} \right) & = & \\
-c(uv) \cdot m_{12} & &
\end{eqnarray*}
\item $v \in X_1$ and $u \in X_2$. In this case when we considered the arc $uv$ above (when defining the $w$-weights) we
let $w(uv)=c(uv) \times \frac{m_{11} + m_{22} - m_{12} - m_{21}}{2}$ and added
$c(uv) \times \left(\frac{-m_{22}}{2}\right)$ to $w(su)$ and
$c(uv) \times \frac{m_{12}-m_{11}-m_{21}}{2}$ to $w(tv)$.
So, we added the following amount to the $w$-weight of the  $(s,t)$-cut, $(X_1,X_2)$.
\begin{eqnarray*}
c(uv) \times \left( \frac{m_{11} + m_{22} - m_{12} - m_{21}}{2} + \frac{-m_{22}}{2} + \frac{m_{12}-m_{11}-m_{21}}{2} \right) & = & \\
-c(uv) \cdot m_{21} & &
\end{eqnarray*}
\end{itemize}

Therefore we note that in all cases we have added $-w^P(uv)$ to the $w$-weight of the  $(s,t)$-cut, $(X_1,X_2)$.
So the total $w$-weight of the  $(s,t)$-cut is $-w^P(D)$ as desired.

Analogously, if we have a partition $P=(X_1,X_2)$ of $V(D)$, then adding $s$ to $X_1$ and $t$ to $X_2$ we obtain a $(s,t)$-cut
with $w$-weight $-w^P(D)$ of $H$. 
As we can find a minimum $w^*$-weight cut in $H$ in polynomial time\footnote{Often the Min Cut problem is formulated for graphs rather than multigraphs (and $H$ is a multigraph), but we can easily reduce 
the Min Cut problem from multigraphs to graphs by merging all edges with the same end-points to one edge with the same end-points and with weight equal the sum of the weights of the merged edges.}
 we can find a  partition $P=(X_1,X_2)$ of $V(D)$ with 
minimum value of $-w^P(D)$, which corresponds to the maximum value of $w^P(G)$. Therefore, 
{\it \mwdp(${\cal F}$)} can be solved in polynomial time in this case.~\qed

\begin{proposition}\label{thm2}
If $m_{11} = \max\{ m_{11}, m_{22}, m_{12}, m_{21} \}$ for all matrices $M \in {\cal F}$ then {\it \mwdp(${\cal F}$)} can be solved in polynomial time.
\end{proposition}

\pf
In this case the maximum is clearly obtained by the partition $P=(X_1,X_2)=(V(D),\emptyset)$ as the maximum possible value of $w^P(a)$ for every $a \in A(D)$
is obtained when both endpoints of $a$ belong to $X_1$.  Therefore, the problem is polynomial time solvable.~\qed


\begin{proposition}\label{thm3}
If $m_{22} = \max\{ m_{11}, m_{22}, m_{12}, m_{21} \}$ for all matrices $M \in {\cal F}$ then {\it \mwdp(${\cal F}$)} can be solved in polynomial time.
\end{proposition}

\pf
In this case the maximum is clearly obtained by the partition $P=(X_1,X_2)=(\emptyset,V(D))$ as the maximum possible value of $w^P(a)$ for every $a \in A(D)$
is obtained when both endpoints of $a$ belong to $X_2$.  Therefore, the problem is polynomial time solvable.~\qed

\section{Proving \NP-Hardness Part of Dichotomy}\label{sec:proof2}

\GG{This proof is partitioned into two parts: Part 1 deals with the case of $|{\cal F}|=1$ (see Section \ref{sec:F1}) and Part 2 with the case of arbitrary $|{\cal F}|$ (see Section \ref{sec:Fany}).}

\subsection{Proving \NP-Hardness When $|{\cal F}|=1$}\label{sec:F1}

A hypergraph, $H=(V,E)$, is $3$-{\em uniform} if all edges of $H$ contain three vertices. 
Two edges {\em overlap} if they have at least two vertices in common.
A hypergraph is {\em linear} if it contains no overlapping edges.
A hypergraph is {\em 2-colorable} if the vertex set can be colored with two colors such that every edge contains vertices of both colors.
The following result is well-known.

\2

\begin{theorem}\label{thmL}\cite{Lovasz}\footnote{In fact,  \cite{Lovasz} proves the result for 4-uniform hypergraphs, but a simple gadget allows one to reduce it to 3-uniform hypergraphs. Also, note that the 2-colorability of 3-uniform hypergraphs problem is equivalent to the monotone NAE 3-SAT problem whose \NP-completeness follows from  Schaefer's dichotomy theorem
\cite{Schaefer78}, where monotonicity means that no negations are allowed in the SAT formulas.} 
It is \NP-hard to decide if a $3$-uniform hypergraph is 2-colorable.
\end{theorem}

\2

We will extend the above theorem to the following theorem, where we only consider linear hypergraphs.

\begin{theorem} \label{thmNPhard}
It is \NP-hard to decide if a linear $3$-uniform hypergraph is 2-colorable.
\end{theorem}

\pf
  We will reduce from the problem to decide if a $3$-uniform hypergraph is 2-colorable, which is \NP-hard by  Theorem \ref{thmL}.
So let $H$ be a $3$-uniform hypergraph. We will reduce $H$ to a linear $3$-uniform hypergraph, $H'$, such that $H$ is 2-colorable if and only if
$H'$ is 2-colorable. This will complete the proof.

We may assume that $H$ is not linear, as otherwise we just let $H'=H$ and we are done. Let $e_y=\{x_1,x_2,y\}$ and $e_z=\{x_1,x_2,z\}$ be two edges that intersect
in at least two vertices. 
Let $F$ be a copy of the Fano plane and let $e=\{f_1,f_2,f_3\}$ be any edge in $F$. It is known that $F$ is not 2-colorable but $F-e$ is 2-colorable and any 
2-coloring of $F-e$ assigns the same color to all vertices of $e$ (as $F$ is not 2-colorable). Now add $F-e$ to $H-x_1$ and for every edge in $H$ that contains
$x_1$ replace $x_1$ by some vertex in $\{f_1,f_2,f_3\}$. Furthermore do this such that $e_y$ and $e_z$ are given different vertices in $\{f_1,f_2,f_3\}$.
Note that the resulting hypergraph is 2-colorable if and only if $H$ is 2-colorable. Furthermore, the number of edges that overlap has decreased. 
So, we have to perform the above operation at most ${|E(H)| \choose 2} \leq |E(H)|^2$ times in order to get rid of all overlapping edges. So this reduction is polynomial
and results in a linear $3$-uniform hypergraph, $H'$, that is 2-colorable if and only if
$H$ 2-colorable.~\qed

\begin{theorem} \label{thm4}
If $m_{11} + m_{22} < m_{12} + m_{21}$ and $\max\{m_{11},m_{22}\} < \max\{ m_{12}, m_{21} \}$  for some matrix $M \in {\cal F}$ then {\it \mwop(${\cal F}$)} is NP-hard to solve.
\end{theorem}
\pf
We will reduce from the problem of determining if a linear $3$-uniform hypergraph is bipartite, which is NP-hard by Theorem~\ref{thmNPhard}.
Let $H=(V,E)$ be a linear $3$-uniform hypergraph. We will now construct an instance of {\it \mwdp(${\cal F}$)}, such that a solution to this instance
will tell us if $H$ is bipartite or not.

We start by letting $V(D)=V(H)$ and for each edge $e \in E(H)$ we add a directed $3$-cycle $C_e=xyzx$ to $D$, where $V(e)=\{x,y,z\}$.
Furthermore for each arc $a$ in the directed $3$-cycle let $c(a)=1$ and $f(a)=M$ ($M$ is defined in the statement of the theorem).
As $H$ is linear we note that $D$ is an oriented graph. 
We now consider the following cases.

\2

{\bf Case 1. $m_{11}=m_{22}$:} In this case we will show that the above construction of $(D,c,f)$ is sufficient.
If $P=(X_1,X_2)$ is a partition on $V(H)$ (and therefore also a partition of $V(D)$) and $e \in E(H)$ then the following holds.

\begin{itemize}
 \item If $|V(e) \cap X_1|=3$ then the arcs of $C_e$ contribute $3m_{11}$ to $w^P(D)$.
 \item If $|V(e) \cap X_1|=2$ then the arcs of $C_e$ contribute $m_{11}+m_{12}+m_{21}$ to $w^P(D)$.
 \item If $|V(e) \cap X_1|=1$ then the arcs of $C_e$ contribute $m_{22}+m_{12}+m_{21}$ to $w^P(D)$.
 \item If $|V(e) \cap X_1|=0$ then the arcs of $C_e$ contribute $3m_{22}$ to $w^P(D)$.
\end{itemize}

So, as $m_{11}=m_{22}$, this implies the following.

\begin{itemize}
 \item If $|V(e) \cap X_1| \in \{0,3\}$ then the arcs of $C_e$ contribute $3m_{11}$ to $w^P(D)$.
 \item If $|V(e) \cap X_1| \in \{1,2\}$ then the arcs of $C_e$ contribute $m_{11}+m_{12}+m_{21}$ to $w^P(D)$.
\end{itemize}

And as $m_{12}+m_{21} > m_{11}+m_{22} = 2m_{11}$, we note that the maximum value of $w^P(D)$ is $|V(H)| (m_{11}+m_{12}+m_{21})$ if and only if
$|V(e) \cap X_1| \in \{1,2\}$ for all $e \in E(H)$, which is equivalent to $H$ being bipartite (as $X_1$ are the vertices with one color and $X_2$
are the vertices with the other color). So in this case we have the desired reduction.

\2

{\bf Case 2. $m_{11} \not= m_{22}$:}  Without loss of generality we may assume that $m_{11} > m_{22}$.
We now construct an instance, $(G(x,y),c,f)$, of {\it \mwop(${\cal F}$)} such that every optimal solution $(X_1,X_2)$ for $G(x,y)$ has $x \in X_1$ and $y \in X_2$.

If $m_{12} > m_{21}$ then $m_{12}$ is the unique largest value in $\{ m_{11}, m_{22}, m_{12}, m_{21} \}$. 
Let $G(x,y)$ have vertex set $\{x,y\}$ and contain a single arc $xy$ with $f(xy)=M$ and $c(xy)=1$. Then
$x \in X_1$ and $y \in X_2$ for all optimal solutions $(X_1,X_2)$ to $G(x,y)$. 

If $m_{21} > m_{12}$ then $m_{21}$ is the unique largest value in $\{ m_{11}, m_{22}, m_{12}, m_{21} \}$. 
Analogously to the above construction, let $G(x,y)$ have vertex set $\{x,y\}$ and contain a single arc $yx$ with $f(yx)=M$ and $c(xy)=1$. Then 
$x \in X_1$ and $y \in X_2$ for all optimal solutions $(X_1,X_2)$ to $G(x,y)$.

Finally consider the case when $m_{12} = m_{21}$. 
Let  $G(x,y)$ have vertex set $\{x,y,z\}$ and contain a directed 
$3$-cycle $xyzx$ with $f(xy)=f(yz)=f(zx)=M$ and $c(xy)=c(yz)=2$ and $c(zx)=1$. Then
$x \in X_1$ (and $z \in X_1$) and $y \in X_2$ for all optimal solutions $(X_1,X_2)$ to $G(x,y)$ (as $m_{12}>m_{11} > m_{22}$).

Let $m^* = \max\{m_{12},m_{21}\}$ and let  $\theta=\frac{m_{11}-m_{22}}{m^*-m_{11}}$. Note that $\theta > 0$ as $m_{11} > m_{22}$ and $m^* > m_{11}$.
We now add $G(x,y)$ to $D$ (defined beforw Case 1) and add arcs between $G(x,y)$ and $D$ as follows.

\begin{itemize}
 \item If $m_{12} > m_{21}$. In this case add all possible arcs, $a$, from $x$ to $V(D)$ and let $c(a)=0$ and $f(a)=M$ for each of these.
Now for each $e \in E(H)$ we add $\theta$ to the $c$-value of the three arcs from $x$ to $V(C_e)$. This completes the construction.

 \item If $m_{12} \leq m_{21}$. In this case add all possible arcs, $a$, from $V(D)$ to $x$ and let the $c(a)=0$ and $f(a)=M$ for each of these.
Now for each $e \in E(H)$ we add $\theta$ to the $c$-value of the three arcs from $V(C_e)$ to $x$. This completes the construction.
\end{itemize}

Let $D^*$ denote the resulting digraph.  Now multiply the $c$-values of all arcs in $G(x,y)$ by a large enough constant to force
$x \in X_1$ and $y \in X_2$ for all optimal solutions $(X_1,X_2)$ to $D^*$.
Let $P=(X_1,X_2)$ be any partition of $V(H)$.  The following now holds for each $e \in E(H)$.

\begin{description}
 \item[If $|V(e) \cap X_1|=3$:] Then the arcs of $C_e$ contribute $3m_{11}$ to $w^P(D^*)$ and the weights we added to the arcs between $x$ and $V(e)$ when 
considering $e$ contribute $3 \theta m_{11}$ to $w^P(D^*)$. So all-in-all $w^P(D^*)$ has increased by the following amount due to $e$,
\[
 s_3 = 3m_{11} + 3 \theta m_{11} 
\]
 \item[If $|V(e) \cap X_1|=2$:] Then the arcs of $C_e$ contribute $m_{11}+m_{12}+m_{21}$ to $w^P(D^*)$ and the weights we added to the arcs between $x$ and $V(e)$ when 
considering $e$ contribute $2 \theta m_{11} + \theta m^*$ to $w^P(D^*)$. So all-in-all $w^P(D^*)$ has increased by the following amount due to $e$,
\[
 s_2 = m_{11}+m_{12}+m_{21} + \theta(2 m_{11} + m^*)
\]
 \item[If $|V(e) \cap X_1|=1$:] Then  the arcs of $C_e$ contribute $m_{22}+m_{12}+m_{21}$ to $w^P(D^*)$ and the weights we added to the arcs between $x$ and $V(e)$ when
considering $e$ contribute $\theta m_{11} + 2 \theta m^*$ to $w^P(D^*)$. So all-in-all $w^P(D^*)$ has increased by the following amount due to $e$,
 \[
 s_1 = m_{22}+m_{12}+m_{21} + \theta(m_{11} + 2 m^*)
\]
\item[If $|V(e) \cap X_1|=0$:] Then  the arcs of $C_e$ contribute $3m_{22}$ to $w^P(D^*)$ and the weights we added to the arcs between $x$ and $V(e)$ when
considering $e$ contribute $3 \theta m^*$ to $w^P(D^*)$. So all-in-all $w^P(D^*)$ has increased by the following amount due to $e$,
 \[
 s_0 = 3m_{22} + 3 \theta m^*
\]
\end{description}

We will now show that $s_0=s_3 < s_2=s_1$, which implies that $H$ is bipartite if and only if the optimal solution, $P=(X_1,X_2)$, for $D^*$ has 
 $|V(e) \cap X_1| \in \{1,2\}$ for all $e \in E(H)$. So we have reduced  
 the problem of determining if a linear $3$-uniform hypergraph is 2-colorable (which is \NP-hard by Theorem~\ref{thmNPhard}) to {\it \mwop(${\cal F}$)}
as desired. So the following claims complete the proof.

\2

{\bf Claim 2.A:} $s_0=s_3$.

{\bf Proof of Claim 2.A:} Note that the following holds as $\theta=\frac{m_{11}-m_{22}}{m^*-m_{11}}$, which proves the claim.
\[
\begin{array}{rcl}
s_3-s_0 & = & (3m_{11} + 3 \theta m_{11})-(3m_{22} + 3 \theta m^*) \\
        & = & 3(m_{11}-m_{22}) + 3 \frac{m_{11}-m_{22}}{m^*-m_{11}} (m_{11} - m^*) \\
        & = & 0 \\
\end{array}
\]
\2

{\bf Claim 2.B:} $s_2=s_1$.

{\bf Proof of Claim 2.B:}  Note that the following holds as $\theta=\frac{m_{11}-m_{22}}{m^*-m_{11}}$, which proves the claim.
\[
s_2-s_1  =  m_{11} - m_{22} + \theta (m_{11} - m^*) = (m_{11} - m_{22})-(m_{11} - m_{22}) = 0
\]
\2

{\bf Claim 2.C:} $s_3 < s_2$.

{\bf Proof of Claim 2.C:} Note that the following holds as $\theta=\frac{m_{11}-m_{22}}{m^*-m_{11}}$, which proves the claim.
\[
\begin{array}{rcl}
s_2-s_3 & = & (m_{12}+m_{21}-2m_{11}) +  \theta (m^* - m_{11}) \\
        & = & (m_{12}+m_{21}-2m_{11}) +  (m_{11}-m_{22}) \\
        & = & (m_{12}+m_{21}) - (m_{11}+m_{22}) \\
        & > & 0 \\
\end{array}
\]
\mbox{ } \hfill \qed

Note that Theorem~\ref{lem:case-1}, Proposition~\ref{thm2}, Proposition~\ref{thm3} and Theorem~\ref{thm4} imply a dichotomy for {\it \mwdp(${\cal F}$)} when 
${\cal F}$ only contains a single matrix. In order to obtain a complete dichotomy for all ${\cal F}$ we need the results in the next subsection.

\subsection{Proving \NP-Hardness for Arbitrary $|{\cal F}|$}\label{sec:Fany}

Let $M$ be a $2 \times 2$ matrix. Let us recall the following three properties of $M$ which may, or may not, hold.

\begin{description}
\item[Property (a):] $m_{11} + m_{22} \geq m_{12} + m_{21}$.
\item[Property (b):] $m_{11} = \max\{ m_{11}, m_{22}, m_{12}, m_{21} \}$.
\item[Property (c):] $m_{22} = \max\{ m_{11}, m_{22}, m_{12}, m_{21} \}$.
\end{description}

Note that Theorem~\ref{lem:case-1}, Proposition~\ref{thm2}, Proposition~\ref{thm3} and Theorem~\ref{thm4} can be reformulated as follows.
\begin{corollary}\label{cor1}
{\it \mwdp(${\cal F}$)} can be solved in polynomial time if all matrices in ${\cal F}$ satisfy Property (a) or all matrices in ${\cal F}$ satisfy Property (b)
or all matrices in ${\cal F}$ satisfy Property (c).  {\it \mwop(${\cal F}$)} is \NP-hard if some matrix in ${\cal F}$ does not satisfy any of the properties (a), (b) or (c).
\end{corollary}


The next two theorems are required for proving Theorem \ref{NPhMWOP}.

\begin{theorem} \label{thm5}
We consider {\it \mwop(${\cal F}$)}, where $M$ and $R$ are distinct matrices in ${\cal F}$.
If $M$ satisfies Property (b), but not Property (a) or Property (c) and $R$ satisfy Property (c), but not Property (b) (it may, or may not, 
satisfies Property (a)) then {\it \mwop(${\cal F}$)} is \NP-hard to solve.

By symmetry, if $M$ satisfies Property (c), but not Property (a) or Property (b) and $R$ satisfies Property (b), but not Property (c) (it may, or may not,
satisfy Property (a)) then the problem is also \NP-hard to solve.
\end{theorem}

\pf
Let $M$ and $R$ be defined as in the statement of the theorem.  That is, the following holds.

\begin{description}
\item[(a):] $m_{11} = \max\{ m_{11}, m_{22}, m_{12}, m_{21} \}$ (as Property (b) holds for $M$).
\item[(b):] $m_{11} + m_{22} < m_{12} + m_{21}$ (as Property (a) does not hold for $M$)
\item[(c):] $m_{22} < \max\{ m_{11}, m_{22}, m_{12}, m_{21} \}$ (as Property (c) does not hold for $M$)
\item[(d):] $r_{22} = \max\{ r_{11}, r_{22}, r_{12}, r_{21} \}$ (as Property (c) holds for $R$). 
\item[(e):] $r_{11} < \max\{ r_{11}, r_{22}, r_{12}, r_{21} \}$ (as Property (b) does not hold for $R$)
\item[(f):] $m_{11} > m_{22}$, as by (a) and (c) we have $m_{11} = \max\{ m_{11}, m_{22}, m_{12}, m_{21} \} > m_{22}$. 
\item[(g):] $r_{22} > r_{11}$, as by (d) and (e) we have $r_{22} = \max\{ r_{11}, r_{22}, r_{12}, r_{21} \} > r_{11}$.
\end{description}

We will first construct an instance  $(D(x,x',y,y'),c,f)$ of {\it \mwdp(${\cal F}$)} such that for all optimal solutions
$P=(X_1,X_2)$ of the instance we have $x,x' \in X_1$ and $y,y' \in X_2$. We consider the following cases.

\2

{\bf Case 1: $m_{11}=m_{12}=m_{21}$ is not true and $r_{22}=r_{12}=r_{21}$ is also not true.} In this case $V(D(x,x',y,y'))=\{x,x',x'',y,y',y''\}$ and
$A(D(x,x',y,y'))=\{xx',x'x'',x''x,yy',y'y'',y''y\}$. Furthermore $c(a)=1$ for all arcs $a \in A(D(x,x',y,y'))$ and $f(xx')=f(x'x'')=f(x''x)=M$ and
$f(yy')=f(y'y'')=f(y''y)=R$. The only optimal solution $P=(X_1,X_2)$ is now $X_1=\{x,x',x''\}$ and $X_2=\{y,y',y''\}$ with value $w^P(D(x,x',y,y'))=3m_{11}+3r_{22}$.

\2

{\bf Case 2: $m_{11}=m_{12}=m_{21}$ and $r_{22}=r_{12}=r_{21}$.}   In this case $V(D(x,x',y,y'))=\{x,x',x_2,x_3,y,y',y_2,y_3\}$ and
$$A(D(x,x',y,y'))=\{xx_2,xx_3,x'x_2,x'x_3,x_2x_3,yy_2,yy_3,y'y_2,y'y_3, y_2y_3\}.$$ Furthermore $c(a)=1$ for all arcs $a \in A(D(x,x',y,y'))$ and 
$f(xx_2)=f(xx_3)=f(x'x_2)=f(x'x_3)=f(y_2y_3)=M$ and
$f(yy_2)=f(yy_3)=f(y'y_2)=f(y'y_3)=f(x_2x_3)=R$. 
We note that $X_1=\{x,x',y_2,y_3\}$ and $X_2=\{y,y',x_2,x_3\}$ is an optimal solution with value $w^P(D(x,x',y,y'))=m_{11}+4m_{12}+r_{22}+4r_{21}
=5m_{11}+5r_{22}$.

For the sake of contradiction assume that some other optimal solution $P=(X_1,X_2)$ has $x \in X_2$. If both $x_2$ and $x_3$ are in $X_1$ then 
$w^P(x_2x_3)=r_{11} < r_{22}$, contradicting the fact that $P$ was optimal.  So either $x_2 \in X_2$ or $x_3 \in X_2$. but in either
case we have an arc $a \in \{xx_2,xx_3\}$ with $w^P(a)=m_{22} < m_{11}$, contradicting the fact that $P$ was optimal. So we must have $x \in X_1$ for all
optimal solutions $P=(X_1,X_2)$.
Analogously we can show that we also must have $x' \in X_1$, $y \in X_2$ and $y' \in X_2$, which completes the proof of Case 2.

\2

{\bf Case 3: $m_{11}=m_{12}=m_{21}$ is not true but $r_{22}=r_{12}=r_{21}$ is true.}  In this case $V(D(x,x',y,y'))=\{x,x',y,y',z\}$ and
$A(D(x,x',y,y'))=\{xx',x'z,zx,yy',y'z,zy\}$. Furthermore $c(a)=1$ for all arcs $a \in A(D(x,x',y,y'))$ and $f(xx')=f(x'z)=f(zx)=M$ and
$f(yy')=f(y'z)=f(zy)=R$.  We note that $X_1=\{x,x',z\}$ and $X_2=\{y,y'\}$ is an optimal solution with value 
$w^P(D(x,x',y,y'))=3m_{11}+r_{22}+r_{12}+r_{21}=3m_{11}+3r_{22}$. It is furthermore not difficult to check that this is the only optimal solution.

\2

{\bf Case 4: $m_{11}=m_{12}=m_{21}$ but $r_{22}=r_{12}=r_{21}$ is not true.}  This can be proved analogously to Case~3.

\2

Now that we have constructed $D(x,x',y,y')$ we can give the NP-hardness proof. We will reduce from the (unweighted) {\sc MaxCut} problem.
Let $G$ be any graph where we wish to find a cut with the maximum number of edges.
 We initially let $V(D^*)=V(G)$. Let $\epsilon=\frac{m_{11}-m_{22}}{r_{22}-r_{11}}$ and note that $\epsilon>0$ by properties~(f) and (g).
For each edge $e \in E(G)$, we now perform the following modifications to $D^*$.

\begin{description}
\item[1.] If $e=uv$ in $G$ then add the arc $uv$ to $D^*$ (any orientation of $e$ can be added to $D^*$) and let $c(uv)=1$ and $f(uv)=M$.
\item[2.] Add a copy, $D_e(x_e,x_e',y_e,y_e')$, of $D(x,x',y,y')$ to $D^*$ and add the following arcs.
\begin{description}
 \item[2a.] Add the directed $4$-cycle $C_e = ux_evy_eu$ to $D^*$ and let $c(a)=\epsilon$ and $f(a)=R$ for all $a \in A(C_e)$.
 \item[2b.] Add the arcs $A_e=\{vx'_e,x'_eu,vy'_e,y'_eu\}$ to $D^*$ and let $c(a)=1/2$ and $f(a)=M$ for all $a \in A(A_e)$.
\end{description}
\item[3.] Multiply $c(a)$ by a large constant, $K$, for all arcs $a$ which are part of a gadget $D_e(x_e,x_e',y_e,y_e')$ such that 
$x_e,x_e' \in X_1$ and $y_e,y_e' \in X_2$ for all optimal solutions $P=(X_1,X_2)$.
\end{description}

Let $P=(X_1,X_2)$ be an optimal solution for the instance  $(D^*,c,f)$. Let $ e=uv \in E(G)$ be arbitrary, such that $uv$ was added 
to $D^*$ in Step~1 above. Now consider the following four cases.
 
\2

{\bf Case A: $u,v \in X_1$.} The arcs added to $D^*$ in Step~1 and Steps~2a and 2b above contribute the following to $w^P(D^*)$.
\[
 s_A = m_{11}+\frac{2m_{11}+m_{12}+m_{21}}{2}+\epsilon ( 2r_{11}+r_{12}+r_{21} ) 
\]
\2
{\bf Case B: $u,v \in X_2$.} The arcs added to $D^*$ in Step~1 and Steps~2a and 2b above contribute the following to $w^P(D^*)$.
\[
 s_B = m_{22}+\frac{2m_{22}+m_{12}+m_{21}}{2}+\epsilon ( 2r_{22}+r_{12}+r_{21} ) 
\]
\2
{\bf Case C: $u \in X_1$ and $v \in X_2$.} The arcs added to $D^*$ in Step~1 and Steps~2a and 2b above contribute the following to $w^P(D^*)$.
\[
 s_C = m_{12}+\frac{2m_{21}+m_{11}+m_{22}}{2}+\epsilon ( r_{11}+r_{22}+r_{12}+r_{21} ) 
\]
\2
{\bf Case D: $u \in X_1$ and $v \in X_2$.} The arcs added to $D^*$ in Step~1 and Steps~2a and 2b above contribute the following to $w^P(D^*)$.
\[
 s_D = m_{21}+\frac{2m_{12}+m_{11}+m_{22}}{2}+\epsilon ( r_{11}+r_{22}+r_{12}+r_{21} )
\]

\2

We immediately see that $s_C=s_D$ as they are both equal to $m_{21}+m_{12} + (m_{11}+m_{22})/2+\epsilon ( r_{11}+r_{22}+r_{12}+r_{21} )$.
We will now show that $s_A=s_B$, which follows from the following, as $\epsilon=\frac{m_{11}-m_{22}}{r_{22}-r_{11}}$.
\[
\begin{array}{crcl}
& m_{11}-m_{22}  
& = &   \epsilon ( r_{22} - r_{11}) \\
\Updownarrow & & & \\
& 2m_{11}  + 2 \epsilon r_{11} 
& = &  2m_{22}+2 \epsilon r_{22} \\
\Updownarrow & & & \\
& m_{11}+\frac{2m_{11}+m_{12}+m_{21}}{2}+\epsilon ( 2r_{11}+r_{12}+r_{21} ) 
& = &  m_{22}+\frac{2m_{22}+m_{12}+m_{21}}{2}\\
&&& +\epsilon ( 2r_{22}+r_{12}+r_{21} ) \\
\Updownarrow & & & \\
& s_A & = & s_B \\ 
\end{array}
\]
This implies that $s_A=s_B$. We also note that the following holds.
\[
\begin{array}{rclcl}
(s_C+s_D)-(s_A+s_B) & = & (m_{12}+m_{21})-(m_{11}+m_{22}) & > & 0 \\
\end{array}
\]
So for every edge $e \in E(G)$ belonging to the cut $(X_1,X_2)$ we add $s_C$ ($=s_D$) to the value $w^P(D^*)$ and 
for every edge  $e \in E(G)$ not belonging to the cut $(X_1,X_2)$ we add $s_A$ ($=s_B<s_C=s_D$). 
So an optimal solution to max-cut for $G$ is equivalent to an optimal solution $(X_1,X_2)$ of $(D^*,c,f)$.
This completes the proof of the first part of the theorem.

The second part follows immediately by symmetry.~\qed

\begin{theorem} \label{thm6}
We consider {\it $\mwop({\fcal}$)}, where $M$ and $R$ are distinct matrices in ${\cal F}$.
If $M$ satisfies Property (b), but not Property (a) or Property (c) and $R$ satisfies Property (a), but not Property (b) or Property (c)
 then {\it $\mwop({\cal F}$)} is \NP-hard to solve.

By symmetry, if $M$ satisfies Property (c), but not Property (a) or Property (b) and $R$ satisfies Property (a), but not Property (b) or Property (c)
 then the problem is also \NP-hard to solve.
\end{theorem}

\pf
Let $M$ and $R$ be defined as in the statement of the theorem.  That is, the following holds.

\begin{description}
\item[(a):] $m_{11} = \max\{ m_{11}, m_{22}, m_{12}, m_{21} \}$ (as Property (b) holds for $M$).
\item[(b):] $m_{11} + m_{22} < m_{12} + m_{21}$ (as Property (a) does not hold for $M$)
\item[(c):] $m_{22} < \max\{ m_{11}, m_{22}, m_{12}, m_{21} \}$ (as Property (c) does not hold for $M$)
\item[(d):] $r_{11} + r_{22} \geq r_{12} + r_{21}$ (as Property (a) holds for $R$).
\item[(e):] $r_{22} < \max\{ r_{11}, r_{22}, r_{12}, r_{21} \}$ (as Property (c) does not hold for $R$).
\item[(f):] $r_{11} < \max\{ r_{11}, r_{22}, r_{12}, r_{21} \}$ (as Property (b) does not hold for $R$)
\item[(g):] $m_{11} > m_{22}$, as by (a) and (c) we have $m_{11} = \max\{ m_{11}, m_{22}, m_{12}, m_{21} \} > m_{22}$.
\item[(h):] $r_{12} \not= r_{21}$, as if $r_{12} = r_{21}$, then $r_{12} = r_{21} = \max\{ r_{11}, r_{22}, r_{12}, r_{21} \}$, by (e) and (f),
which is impossible by (d) (and (e) and (f)).
\end{description}

We will first construct an instance $(D(x,x',y,y'),c,f)$ of {\it \mwop(${\cal F}$)} such that for all optimal solutions
$P=(X_1,X_2)$ of the instance we have $x,x' \in X_1$ and $y,y' \in X_2$. We consider the following cases, which exhaust all
possiblities by statement~(h) above.

\2

{\bf Case 1: $r_{12}>r_{21}$.} By (e) and (f) we note that $r_{12}$ is the unique largest value in $\{r_{11}, r_{22}, r_{12}, r_{21}\}$. 
Let $V(D(x,x',y,y'))=\{x,x',y,y'\}$ and $A(D(x,x',y,y'))=\{xy,x'y'\}$. 
Furthermore $c(xy)=c(x'y')=1$ and $f(xy)=f(x'y')=R$.
The only optimal solution $P=(X_1,X_2)$ is now $X_1=\{x,x'\}$ and $X_2=\{y,y'\}$ with value $w^P(D(x,x',y,y'))=2r_{12}$.

\2

{\bf Case 2: $r_{21}>r_{12}$.} By (e) and (f) we note that $r_{21}$ is the unique largest value in $\{r_{11}, r_{22}, r_{12}, r_{21}\}$. 
Let $V(D(x,x',y,y'))=\{x,x',y,y'\}$ and $A(D(x,x',y,y'))=\{yx,y'x'\}$. 
Furthermore $c(yx)=c(y'x')=1$ and $f(yx)=f(y'x')=R$.
The only optimal solution $P=(X_1,X_2)$ is now $X_1=\{x,x'\}$ and $X_2=\{y,y'\}$ with value $w^P(D(x,x',y,y'))=2r_{21}$.

\2

Now that we have constructed $D(x,x',y,y')$ we can give the \NP-hardness proof. We will reduce from the (unweighted) {\sc MaxCut} problem.
So let $G$ be any graph where we are interested in finding a cut with the maximum number of edges.
 We initially let $V(D^*)=V(G)$.  Let $r^* = \max \{r_{12},r_{21}\}$ and let $\epsilon=\frac{m_{11}-m_{22}}{r^*-r_{11}}$ and note that $\epsilon>0$ by 
properties~(g), (e) and (f).
For each edge $e \in E(G)$ we now perform the following modifications to $D^*$.

\begin{description}
\item[1.] If $e=uv$ in $G$ then add the arc $uv$ to $D^*$ (any orientation of $e$ can be added to $D^*$) and let $c(uv)=1$ and $f(uv)=M$.
\item[2.] Add a copy, $D_e(x_e,x_e',y_e,y_e')$, of $D(x,x',y,y')$ to $D^*$ and add the following arcs.
\begin{description}
 \item[2a.] If $r_{12}>r_{21}$ then add the arcs $A_e'=\{x_eu,x_ev\}$ to $D^*$ and let $c(x_eu)=c(x_ev)=\epsilon$ and $f(x_eu)=f(x_ev)=R$.

And if $r_{12}<r_{21}$ then add the arcs $A_e'=\{ux_e,vx_e\}$ to $D^*$ and let $c(u_ex)=c(vx_e)=\epsilon$ and $f(ux_e)=f(vx_e)=R$.

 \item[2b.] Add the arcs $A_e=\{vx'_e,x'_eu,vy'_e,y'_eu\}$ to $D^*$ and let $c(a)=1/2$ and $f(a)=M$ for all $a \in A(A_e)$.
\end{description}
\item[3.] Multiply $c(a)$ by a large constant, $K$, for all arcs $a$ which are part of a gadget $D_e(x_e,x_e',y_e,y_e')$ such that 
$x_e,x_e' \in X_1$ and $y_e,y_e' \in X_2$ for all optimal solutions $P=(X_1,X_2)$.
\end{description}

Let $P=(X_1,X_2)$ be an optimal solution for the instance  $(D^*,c,f)$. Let $e=uv \in E(G)$ be arbitrary, such that $uv$ was added 
to $D^*$ in Step~1 above. Now consider the following four cases.
 
\2

{\bf Case A: $u,v \in X_1$.} The arcs added to $D^*$ in Step~1 and Steps~2a and 2b above contribute the following to $w^P(D^*)$.
\[
 s_A = m_{11}+\frac{2m_{11}+m_{12}+m_{21}}{2}+\epsilon (2r_{11}) 
\]
\2
{\bf Case B: $u,v \in X_2$.} The arcs added to $D^*$ in Step~1 and Steps~2a and 2b above contribute the following to $w^P(D^*)$.
\[
 s_B = m_{22}+\frac{2m_{22}+m_{12}+m_{21}}{2}+\epsilon ( 2r^* ) 
\]
\2
{\bf Case C: $u \in X_1$ and $v \in X_2$.} The arcs added to $D^*$ in Step~1 and Steps~2a and 2b above contribute the following to $w^P(D^*)$.
\[
 s_C = m_{12}+\frac{2m_{21}+m_{11}+m_{22}}{2}+\epsilon ( r^*+r_{11} )
\]
\2
{\bf Case D: $u \in X_2$ and $v \in X_1$.} The arcs added to $D^*$ in Step~1 and Steps~2a and 2b above contribute the following to $w^P(D^*)$.
\[
 s_D = m_{21}+\frac{2m_{12}+m_{11}+m_{22}}{2}+\epsilon ( r^*+r_{11} )
\]
\2

We immediately see that $s_C=s_D$ as they are both equal to $m_{21}+m_{12} + (m_{11}+m_{22})/2+ \epsilon ( r^*+r_{11} )$.
We will now show that $s_A=s_B$, which follows from the following, as $\epsilon=\frac{m_{11}-m_{22}}{r^*-r_{11}}$.

\[
\begin{array}{crcl}
& m_{11}-m_{22}  
& = &   \epsilon ( r^* - r_{11}) \\
\Updownarrow & & & \\
& 2m_{11}  + 2 \epsilon r_{11} 
& = &  2m_{22}+2 \epsilon r^* \\
\Updownarrow & & & \\
& m_{11}+\frac{2m_{11}+m_{12}+m_{21}}{2}+ 2\epsilon r_{11}  
& = &  m_{22}+\frac{2m_{22}+m_{12}+m_{21}}{2}+2\epsilon  r^* \\
\Updownarrow & & & \\
& s_A & = & s_B \\ 
\end{array}
\]

This implies that $s_A=s_B$. We also note that the following holds.

\[
\begin{array}{rclcl}
(s_C+s_D)-(s_A+s_B) & = & (m_{12}+m_{21})-(m_{11}+m_{22}) & > & 0 \\
\end{array}
\]

So for every edge $e \in E(G)$ belonging to the cut $(X_1,X_2)$ we add $s_C$ ($=s_D$) to the value $w^P(D^*)$ and 
for every edge  $e \in E(G)$ not belonging to the cut $(X_1,X_2)$ we add $s_A$ ($=s_B<s_C=s_D$). 
So an optimal solution to max-cut for $G$ is equivalent to an optimal solution $(X_1,X_2)$ of $(D^*,c,f)$.
This completes the proof of the first part of the theorem.

The second part follows immediately by symmetry.~\qed



\2

\begin{theorem}\label{NPhMWOP}
If not all matrices in ${\cal F}$ satisfy Property (a) and not all matrices in ${\cal F}$ satisfy Property (b) and not all matrices in ${\cal F}$ satisfy Property (c), then
\mwop(${\cal F}$)  is \NP-hard. 
\end{theorem}

\pf
Assume that  some matrix in ${\cal F}$ does not satisfy Property (a), some matrix (possibly the same)
does not satisfy Property (b) and some matrix (possibly the same)
does not satisfy Property (c).
Let $M$ be a matrix in ${\cal F}$ that does not satisfy Property (a). If $M$ does not satisfy Property (b) and does not satisfy Property (c) then 
we are done by Theorem~\ref{thm4}, so, without loss of generality, we may assume that $M$ satisfies Property (b). If $M$ satisfies
Property (c), then $m_{11} = m_{22} = \max\{ m_{11}, m_{22}, m_{12}, m_{21} \}$, which implies that Property (a) holds, a contradiction.
So assume that $M$ satisfies Property (b), but not Property (a) or Property (c).

Let $R$ be a matrix in ${\cal F}$ that does not satisfy Property (b). 
If $R$ satisfies Property (c), then we are done by Theorem~\ref{thm5}, so we may assume that this is not the case.
If $R$ satisfies Property (a), then we are done by Theorem~\ref{thm6}, so we may assume that this is not the case.
But now $R$ does not satisfy Property (a), Property (b) or Property (c) and we are done by Theorem~\ref{thm4}.
\qed

\GG{
\begin{theorem}\label{NPhMWSDP}
If not all matrices in ${\cal F}$ satisfy Property (a) and not all matrices in ${\cal F}$ satisfy Property (b) and not all matrices in ${\cal F}$ satisfy Property (c), then
\mwsdp(${\cal F}$)  is NP-hard. 
\end{theorem}
\pf 
Consider a set $\cal F$ of real 2$\times$2 matrices and 
assume that  some matrix in ${\cal F}$ does not satisfy Property (a), some matrix (possibly the same)
does not satisfy Property (b) and some matrix (possibly the same)
does not satisfy Property (c). Consider the following reduction from $\mwop({\fcal})$ to $\mwsdp({\fcal})$: 
an instance $(D,c,f)\in \mwop({\fcal})$ is mapped to $(D',c',f')\in \mwsdp({\fcal})$ such that $D'$ is obtained from $D$ by adding the opposite arc $ji$ to $D'$ for every arc $ij$ of $D$ and setting 
$c'(ij)=c(ij)$, $c'(ji)=0$, $f'(ij)=f(ij)$ and $f'(ji)=f(ij)$. Note that for every partition $P=(X_1,X_2)$ of $V(D)=V(D')$,
$w^P(D)=w^P(D')$. Thus, \NP-hardness of $\mwop({\fcal})$ implies \NP-hardness of $\mwsdp({\fcal}')$. 
\qed
}

\section{Discussion}\label{sec:disc}

\GG{The main result of this paper is a common computational complexity dichotomy for \mwdp on arbitrary digraphs, oriented graphs and symmetric digraphs. 
We show applications of the dichotomy in game theory and graph theory. Due to applications in  game theory it would be interesting to obtain a dichotomy 
for MWDP on bipartite symmetric digraphs. Since there are $k$-nary-action polymatrix games, the following generalization of MWDP is of interest: instead of  2$\times$2 matrices
we consider $k\times k$ matrices ($k\ge 2$) and instead of bipartitions of $V(D)$, we consider $k$-partitions of  $V(D)$ with the aim of finding a $k$-partition of maximum weight. 
It would be interesting to extend our dichotomy from $k=2$ to arbitrary $k\ge 2.$ }

\paragraph*{Acknowledgements.}
Anders Yeo's research was supported by the Danish research council for independent research under grant number DFF 7014-00037B.



\end{document}